\documentclass[prx, preprint, superscriptaddress, endfloats*]{revtex4-2}
\usepackage{amsmath,hyperref,cleveref,amssymb,amsfonts,graphicx,multirow,color,bm,textcomp,mathtools}
\usepackage{paralist}
\usepackage{srcltx}
\usepackage[all,cmtip]{xy}
\usepackage{mathrsfs}
\usepackage{srcltx,soul,subfigure}
\usepackage{threeparttable,dsfont,bbm}
\usepackage{color}
\usepackage{comment}
\usepackage{dcolumn}

\begin{document}
\title{Observation of 2D Weyl Fermion States in Epitaxial Bismuthene}
\author{Qiangsheng Lu}
\affiliation {Department of Physics and Astronomy, University of Missouri, Columbia, Missouri 65211, USA}
\affiliation {Materials Science and Technology Division, Oak Ridge National Laboratory, Oak Ridge, Tennessee 37831 USA}
\author{P.~V.~Sreenivasa~Reddy}
\affiliation {Department of Physics, National Cheng Kung University, Tainan 701, Taiwan}
\author{Hoyeon Jeon} 
\affiliation {Center for Nanophase Materials Sciences, Oak Ridge National Laboratory, Oak Ridge, Tennessee 37831, USA}
\author{Alessandro R. Mazza} 
\affiliation {Materials Science and Technology Division, Oak Ridge National Laboratory, Oak Ridge, Tennessee 37831 USA}
\affiliation {Center for Integrated Nanotechnologies, Los Alamos National Laboratory, Los Alamos, New Mexico 87123 USA}
\author{Matthew Brahlek}
\affiliation {Materials Science and Technology Division, Oak Ridge National Laboratory, Oak Ridge, Tennessee 37831 USA}
\author{Weikang~Wu}
\affiliation {Research Laboratory for Quantum Materials, Singapore University of Technology and Design, Singapore 487372, Singapore}
\author{Shengyuan A. Yang}
\affiliation {Research Laboratory for Quantum Materials, Singapore University of Technology and Design, Singapore 487372, Singapore}
\author{Jacob~Cook}
\affiliation {Department of Physics and Astronomy, University of Missouri, Columbia, Missouri 65211, USA}
\author{Clayton~Conner}
\affiliation {Department of Physics and Astronomy, University of Missouri, Columbia, Missouri 65211, USA}
\author{Xiaoqian~Zhang}
\affiliation {Department of Physics and Astronomy, University of Missouri, Columbia, Missouri 65211, USA}
\author{Amarnath~Chakraborty}
\affiliation {Department of Physics and Astronomy, University of Missouri, Columbia, Missouri 65211, USA}
\author{Yueh-Ting Yao}
\affiliation {Department of Physics, National Cheng Kung University, Tainan 701, Taiwan}
\author{Hung-Ju Tien}
\affiliation {Department of Physics, National Cheng Kung University, Tainan 701, Taiwan}
\author{Chun-Han Tseng}
\affiliation {Department of Physics, National Cheng Kung University, Tainan 701, Taiwan}
\author{Po-Yuan Yang}
\affiliation {Department of Physics, National Cheng Kung University, Tainan 701, Taiwan}
\author{Shang-Wei Lien}
\affiliation {Department of Physics, National Cheng Kung University, Tainan 701, Taiwan}
\author{Hsin Lin}
\affiliation {Institute of Physics, Academia Sinica, Taipei 11529, Taiwan}
\author{Tai-Chang~Chiang}
\affiliation {Department of Physics, University of Illinois at Urbana-Champaign, 1110 West Green Street, Urbana, Illinois 61801-3080, USA}
\affiliation {Frederick Seitz Materials Research Laboratory, University of Illinois at Urbana-Champaign, 104 South Goodwin Avenue, Urbana, Illinois 61801-2902, USA}
\author{Giovanni~Vignale}
\affiliation {Department of Physics and Astronomy, University of Missouri, Columbia, Missouri 65211, USA}
\author{An-Ping~Li}
\affiliation {Center for Nanophase Materials Sciences, Oak Ridge National Laboratory, Oak Ridge, Tennessee 37831, USA}
\author{Tay-Rong~Chang}
\affiliation {Department of Physics, National Cheng Kung University, Tainan 701, Taiwan}
\affiliation {Center for Quantum Frontiers of Research and Technology (QFort), Tainan 70101, Taiwan}
\affiliation {Physics Division, National Center for Theoretical Sciences, Taipei 10617, Taiwan}
\author{Rob G. Moore}
\affiliation {Materials Science and Technology Division, Oak Ridge National Laboratory, Oak Ridge, Tennessee 37831 USA}
\author{Guang~Bian}
\affiliation {Department of Physics and Astronomy, University of Missouri, Columbia, Missouri 65211, USA}

\newpage
\begin{abstract}
{\bf A two-dimensional (2D) Weyl semimetal featuring a spin-polarized linear band dispersion and a nodal Fermi surface is a new topological phase of matter. It is a solid-state realization of Weyl fermions in an intrinsic 2D system. The nontrivial topology of 2D Weyl cones guarantees the existence of a new form of topologically protected boundary states, Fermi string edge states. In this work, we report the realization of a 2D Weyl semimetal in monolayer-thick epitaxial bismuthene grown on SnS(Se) substrate. The intrinsic band gap of bismuthene is eliminated by the space-inversion-symmetry-breaking substrate perturbations, resulting in a gapless spin-polarized Weyl band dispersion. The linear dispersion and spin polarization of the Weyl fermion states are observed in our spin and angle-resolved photoemission measurements. In addition, the scanning tunneling microscopy/spectroscopy reveals pronounced local density of states at the edge, suggesting the existence of Fermi string edge states. These results open the door for the experimental exploration of the exotic properties of Weyl fermion states in reduced dimensions.}

\end{abstract}


\maketitle

\newpage

\section{Introduction}

The discovery of Dirac materials, which feature massless states near the Fermi level, has driven the rise of the topological era in condensed matter physics. In particular, graphene, a two-dimensional (2D) Dirac semimetal, has enabled the exploration of relativistic physics in tabletop experiments\cite{Novoselov2005, Zhang2005, CastroNeto2009}. The fundamental works on graphene set the foundation for future research ranging from topological insulators, valleytronics, to twistronics\cite{Kane2010, RevModPhys.83.1057, doi:10.1126/science.aao5989, doi:10.1126/science.aat6981, Cao2018-2, Cao2018}. The subsequent generalization from 2D  to 3D platforms prompted the discovery of bulk Dirac semimetals. This generalization has shed light on the behaviors of higher dimensional massless carriers, tilted Dirac cones, and protected surface states\cite{Na3Bi2014, doi:10.1126/science.1256742, PhysRevLett.113.027603, Yan2017, Xiong2015, RevModPhys.90.015001}. The discovery of Weyl semimetals, which host spin-split massless 3D quasiparticles, is particularly exciting since it is the first experimental realization of the Weyl fermion which was proposed long ago in the context of particle physics\cite{PhysRevB.83.205101, Xu_Weyl,PhysRevX.5.031013, Soluyanov2015,RevModPhys.90.015001}. The chiral nodal points and 2D Fermi arc surface states of 3D Weyl semimetals bring about exotic properties such as chiral anomaly, unusual optical conductivity and nonlocal transport\cite{PhysRevB.88.104412, PhysRevX.5.031023, Zhang2016, PhysRevB.87.235306, PhysRevLett.109.196403, PhysRevB.87.245131, Jia2016, PhysRevB.86.115133, PhysRevX.4.031035,PhysRevB.98.035121, Hasan2021}.  Generalization of a 3D Weyl fermion state to 2D gives a new topological state of matter, 2D Weyl semimetals that feature a spin-polarized linear band dispersion and a nodal Fermi surface. A simplistic illustration of Dirac and Weyl semimetals is presented in Fig.~1{\bf a} highlighting the exotic states enabled by reducing the dimensionality from 3D to 2D.  

The 2D Weyl semimetal state was discussed in Haldane's seminal work as a critical phase in the model of quantum anomalous Hall effects \cite{Haldane}. The gapless linear dispersion without spin degeneracy establishes a solid-state realization of Weyl fermions in 2D space. The 2D Weyl fermion states, which correspond to linear bands around Weyl nodes at zero energy, are predicted to exhibit the parity anomaly in (2+1)-D (space–time) quantum field theory \cite{Jackiw, PhysRevLett.57.2967, PhysRevLett.53.2449, Mogi2022} and electron fractionalization with zero modes of charge $e$/2 \cite{PhysRevLett.98.186809}. The fractional zero modes are similar to those in the 1D systems by Jackiw and Rebbi \cite{PhysRevD.13.3398} and by Su, Schrieffer and Hegger \cite{PhysRevLett.42.1698}, and in $p$-wave superconductors by Reed and Green \cite{PhysRevB.61.10267}. The parity anomaly and charge fractionalization are only accessible in 2D Weyl semimetals rather than 2D Dirac semimetals for the quantum numbers would be ``doubled" in spin degenerate states \cite{PhysRevLett.98.186809}. Despite the absence of a band gap, a 2D Weyl semimetal is characterized by a topological invariant, the winding number of the Weyl cone. The winding number can be obtained by integrating the Berry phase along a loop $\ell$ encircling each Weyl node,
\begin{equation}
\theta_\textrm{W} = \oint_{\ell} \bm{A}_{\bm{k}}\cdot {\rm{d}}\bm{k} = \pm \pi,
\end{equation}
where $\bm{A}_{\bm{k}}$ is the berry connection of the occupied valence bands \cite{Lu2016}. The winding number of $\pi$ can be considered as the topological charge of 2D Weyl semimetals, which guarantees the existence of topologically protected edge states \cite{Bian2016, PhysRevB.105.075403}. These topological edge states take the form of Fermi strings (open strings) connecting the projection of bulk Weyl nodes at Fermi level in the edge Brillouin zone as shown Fig.~1{\bf a}. The Fermi string edge states are the 1D analog of Fermi arc surface states in 3D Weyl semimetals. In this sense, 2D Weyl semimetals present a new example of bulk-boundary correspondence in a topological phase. Besides the exotic relativistic and topological properties, the spin and valley degrees of freedom are intrinsically entangled in 2D Weyl semimetals due to the spin polarization of massless electronic states. This unique spin-valley locking behavior of 2D Weyl semimetals gives rise to prominent spin Hall effects and valley Hall effects \cite{Tsai2013}. In addition, the spin-polarized Weyl cones cannot occur around time-reversal invariant momentum (TRIM) points in nonmagnetic 2D materials. Generally, the nodal points are located at generic $\bm{k}$ points with reduced local symmetry, leading to a large asymmetry in the band dispersion. The asymmetric 2D Weyl cones give rise to a dramatic change of Berry curvature dipole (BCD) around the Fermi level, for the Berry curvature density diverges at the nodal point of linear bands \cite{PhysRevLett.115.216806, Du2021} (see Supplementary Information). Therefore, 2D Weyl semimetals provide an ideal platform for studying various nonlinear transport phenomena \cite{PhysRevLett.115.216806}. Those highly unusual properties (sketched in Fig.~1{\bf b}) have inspired a myriad of recent theoretical works on 2D Weyl semimetals \cite{PhysRevResearch.4.043183, Panigrahi2022, PhysRevB.105.075403, PhysRevB.103.L201115, PhysRevB.106.125404}. So far, however, experimental realization of topological 2D Weyl semimetals has remained elusive. 


In this work, we report the discovery of a 2D Weyl semimetal in epitaxial bismuthene (a single atomic layer of bismuth stabilized in phosphorene structure\cite{Liu2014, Li2014}). The 2D Weyl fermion states are achieved through an experimental procedure as shown in Fig.~1{\bf c}. The first step is to find a 2D Dirac material with a narrow SOC gap such as silicene, antimonene, and bismuthene (Fig.~1{\bf c}i)\cite{PhysRevLett.107.076802, PhysRevLett.109.056804, Lu2022, bismuthene2017, Kowalczyk2020, Lu2016}. After a suitable material is found, a perturbation is then applied to break space-inversion symmetry (Fig.~1{\bf c}ii). Such perturbations is commonly found in epitaxial films due to the electric potential from the substrate surface. Like a Rashba system, this perturbation induces spin splitting in the spin-degenerate bands and narrows down the band gap. The spin-split band structure is energetically favored compared to the spin-degenerate case since the energy gain in the down-moving occupied band is larger than the energy cost in the up-moving occupied bands (see Supplementary Information). In other words, the system tends to relax towards a configuration with a larger spin splitting and a narrower gap. With a sufficiently strong substrate perturbation, the band gap can be eliminated, resulting in a Weyl band structure (Fig.~1{\bf c}iii). Using spin- and angle-resolved photoemission spectroscopy (spin-ARPES) and scanning tunneling spectroscopy (STS), we systematically study the electronic structure and spin texture of bismuthene (Bi monolayer in phosphorene structure) grown on SnS and SnSe substrates. The ARPES results demonstrate the spin-polarized gapless Weyl fermion states. Furthermore, the tunneling spectrum shows an enhanced local density of states at the edge of bismuthene, consistent with the calculated edge spectrum with Fermi string states. The experimental results unambiguously establish epitaxial bismuthene on SnS(Se) as an ideal 2D Weyl semimetal. 

\section{The material base for hosting 2D Weyl fermion states}

Bi, a group-Va pnictogen atom, typically forms three covalent bonds with its neighbors. In the 2D limit, two allotropic structural phases, the orthorhombic phosphorene-like phase \cite{Kowalczyk2020} and the hexagonal honeycomb-like phase \cite{science.aai8142}, are allowed by this requirement. The two phases are referred to as $\alpha$- and $\beta$-phases, respectively, in the literature. Here we focus on the phosphorene-like $\alpha$-bismuthene (bismuthene for short in the following discussion). In our experiment, bismuthene thin films were grown with molecular beam epitaxy (MBE). The SnS(Se) was chosen as the substrate as it is van der Waals semiconductor and the (001) surface of SnS(Se) has a similar lattice structure as bismuthene. Figure~2{\bf a} schematically shows the lattice structure of the sample. Bi atoms form a single-layer phosphorene structure on the (001) surface of SnS(Se). The Bi atoms have strong $sp^{3}$-hybridization character and thus the three Bi$-$Bi bonds are close to a tetrahedral configuration. This leads to two atomic sublayers (marked with red and orange colours.) The lattice of bismuthene belongs to the $Pmna$ space group (No.~53). The unit cell has a four-atom basis, which are labeled as A$_\textrm{U}$, A$_\textrm{L}$, B$_\textrm{U}$, and B$_\textrm{L}$ (see the top view in Fig.~2{\bf a}), where U and L refer to the upper and lower sublayer, respectively. The (001)-surface of SnS(Se) has a similar lattice as bismuthene but with S(Se) on A sites and Sn on B sites. The difference in electronegativity of Sn and S(Se) results in an in-plane dipole electric field, which causes an energy shift of A sites relative to B sites on average. In addition, the surface potential of SnS(Se) leads to an electric field perpendicular to the surface and causes a potential difference between the two Bi sublayers. Both the dipole field and the vertical field of the substrate surface break the space-inversion symmetry and give rise to spin splitting in the bands of bismuthene.  We performed scanning-tunneling-microscope (STM) measurements to map the surface topography of the Bi sample grown on SnS(Se). The result is shown in Fig.~2{\bf b}. The observed flat and uniform surface indicates the high structural quality of bismuthene. The red squares marked in the zoom-in STM images (Fig.~2{\bf c}) demonstrate the surface unit cell of bismuthene and SnS. We found the in-plane lattice constants ($a$, $b$) are ($4.5\textrm{\AA}$, $4.8\textrm{\AA}$) for bismuthene, ($4.1\textrm{\AA}$, $4.5\textrm{\AA}$) for SnS surface, and ($4.3\textrm{\AA}$, $4.6\textrm{\AA}$) for SnSe surface, respectively. The apparent height of bismuthene (including the thickness of bimuthene and the spacing between bismuthene and SnS(001) surface) is 8.0$\textrm{\AA}$, which can be seen from the height profile (Fig.~2{\bf d}) taken along the red arrow marked in Fig.~2{\bf b}.

\section{Band structure of the 2D Weyl semimetal} 
Here we present the first-principles band structures and angle-resolved photoemission spectroscopy (ARPES) spectra to demonstrate the existence of the 2D Weyl states in the epitaxial bismuthene films. The calculated band structure of free-standing bismuthene is shown in Fig.~2{\bf e}. The only prominent band feature near the Fermi level is a gapped Dirac cone located at a generic $k$ point between $\overline{{\Gamma}}$ and $\overline{\textrm{X}}_1$. The Dirac bands are dominated by $p_z$ orbitals of Bi atoms. The effective $\bm{k}\cdot\bm{p}$ model around the Dirac point can be written as 

\begin{equation}
\mathcal{H}_{0}^{\text{Dirac}}(\bm{k})=\tau_{\pm}(v_{x}k_{x}\sigma_{x}+\Delta k_{x})+v_{y}k_{y}\sigma_{y}+\tau_{\pm}\lambda_{\rm{SOC}}\sigma_{z}{s_z},
\end{equation}
where ($k_x$, $k_y$) are measured from the Dirac nodes at ($\pm k_0$, 0), $\sigma_i$ ($i=x$,$y$,$z$) are Pauli matrices with respect to the basis of $\{|{\rm A},p_{z}\rangle,|{\rm B},p_{z}\rangle\}$ (the $p_z$ orbitals on the two sublattices A and B), $s_i$ ($i=x$,$y$,$z$) are the spin matrices, $\Delta$ causes a tilting of the Weyl cone in $k_x$ direction and is crucial for generating a non-zero Berry curvature dipole (BCD) \cite{PhysRevLett.115.216806}, $\lambda_{\rm{SOC}}$ is the effective spin-orbit coupling, $\tau_{\pm}=\pm 1$ represents the chirality of the Dirac nodes located at ($\pm k_0$, 0), and $v_{x,y}$ are Fermi velocity along $k_{x}$ and $k_{y}$ directions, respectively.  $v_{x}=3.17\times 10^5$~m/s, $\Delta=0.19\times 10^5$~m/s, $v_{y}=4.23\times 10^5$~m/s, $\lambda_{\rm{SOC}}=55$ meV according to the first-principles results. The energy gap induced by spin-orbit coupling is $\Delta E=2\lambda_{\rm{SOC}}=0.11$~eV. Every band is doubly degenerate with respect to the spin degree of freedom since the lattice of bimuthene is centrosymmetric. The calculated band structure of bismuthene on SnSe is presented in Fig.~2{\bf f}. The presence of SnSe substrate breaks space-inversion symmetry and causes spin splitting in the bands of bismuthene. Particularly, a linear band crossing formed in the same way as the 2D Weyl cone described in Fig.~1{\bf c}. The states from SnSe can be found only below -0.9 eV and above 0.4 eV, because SnSe is a semiconductor with a gap of $\sim$1.3 eV. The 2D Weyl fermion states near the Fermi level are predominantly constructed by Bi $p_z$ orbitals and thus spatially confined in the epitaxial Bi layer, making the system an intrinsic 2D system.

With the inclusion of the substrate, the low-energy effective Hamiltonian is 
\begin{eqnarray}
\mathcal{H}_{0}^{\text{Weyl}}(\bm{k})&=&\mathcal{H}_{0}^{\text{Dirac}}(\bm{k})+\lambda_{\rm Dip}\sigma_{z}+\lambda_{V}'\sigma_{y}s_x+\tau_{\pm}\lambda_{V}''\sigma_{x}s_y \\
&=&\tau_{\pm}(v_{x}k_{x}\sigma_{x}+\Delta k_{x})+v_{y}k_{y}\sigma_{y}+\tau_{\pm}\lambda_{\rm{SOC}}\sigma_{z}{s_z}+\lambda_{\rm Dip}\sigma_{z}+\lambda_{V}'\sigma_{y}s_x+\tau_{\pm}\lambda_{V}''\sigma_{x}s_y,
\end{eqnarray}
where $\lambda_{\rm{Dip}}$ describes the perturbation induced by the in-plane dipole field of the SnSe surface, and $\lambda_{\rm{V}}'$ and $\lambda_{\rm{V}}''$ are the Rashba couplings induced by the vertical electric field from the SnSe surface. The energy gap at the band crossing is 
\begin{equation}
\Delta E=\left|2\lambda_{\rm{SOC}}-\left(\sqrt{\lambda_{\rm Dip}^2+(\lambda_{V}'+\lambda_{V}'')^2}+\sqrt{\lambda_{\rm Dip}^2+(\lambda_{V}'-\lambda_{V}'')^2}\right)\right|. 
\end{equation}
 The quantity $\lambda_{\textrm{Sub}}\equiv\frac{1}{2}\left(\sqrt{\lambda_{\rm Dip}^2+(\lambda_{V}'+\lambda_{V}'')^2}+\sqrt{\lambda_{\rm Dip}^2+(\lambda_{V}'-\lambda_{V}'')^2}\right)$ reflects the strength of substrate effects on the bands of bismuthene. The band gap vanishes when $\lambda_{\rm{SOC}}=\lambda_{\textrm{Sub}}$ (see Supplementary Information). Remarkably, the gapless band dispersion is found in the first-principles band calculation. The calculated 2D Weyl band structure is confirmed by our ARPES measurements. The ARPES results taken from the bismuthene/SnSe sample are plotted in Fig.~3. The Fermi surface (Fig.~3{\bf a}) contains two circular electron pockets in the direction of $\overline{\textrm{X}}_{1}-\overline{{\Gamma}}-\overline{\textrm{X}}_{1}$. We note that a similar pair of electron pockets shows up in the direction of $\overline{\textrm{X}}_{2}-\overline{{\Gamma}}-\overline{\textrm{X}}_{2}$ but with much lower intensity. This extra pair of pockets is due to the existence of Bi domains rotated by $90^{\circ}$ in the MBE sample. The ARPES spectra for bismuthene on SnSe taken along the lines of ``cut1" and ``cut2" (marked in Fig.~3{\bf a}) are plotted in Figs.~3{\bf b-g}. In the ARPES spectra, we found the band dispersion along ``cut1" (``cut2") as well as that from a rotated Bi domain. This can be better seen in Figs.~3{\bf c,f} with overlay of calculated bands on top of the ARPES spectra. The magenta lines are bands along ``cut1" (``cut2") while the green lines are bands along a direction perpendicular to ``cut1" (``cut2"). Only the gapless Weyl cones stay close to the Fermi level, which means the transport and optical properties of this system are entirely determined by the low-energy Weyl fermion states. No apparent gap was found at the nodal point, as evidenced by the second derivative spectrum (Fig.~3{\bf k}) and the map of energy distribution curves (Fig.~3{\bf l}). We also notice that one linear band of Weyl cone is much dimmer than the other in the spectrum of ``cut1". This can be attributed to the photoemission matrix element effects. Figures~3{\bf d, g} shows the calculated spectra with the inclusion of the photoemission matrix elements, which agrees well with the ARPES result. The bismuthene/SnSe sample is electron-doped as the nodal point lies 0.1 eV below the Fermi level. The shift of the Fermi level can be attributed to the electron transfer from the SnSe substrate to the epitaxial bismuthene. We also performed ARPES measurements on bismuthene grown on SnS, and the results are plotted in Figs.~3{\bf h-j}. Compared with bismuthene/SnSe, the Fermi level of bimuthene/SnS is slightly lower (due to the difference in electronegativity of Sn and S) and lies right at the Weyl node. Therefore, bismuthene/SnS is a perfect 2D Weyl semimetal with charge neutrality. Considering the different surface conditions of SnSe and SnS, the observation of Weyl cones in both sample configurations indicate the robustness of the 2D Weyl fermion states against weak perturbations.

\section{Spin texture of the 2D Weyl cone}
The defining character of Weyl fermion states is the spin polarization of the relativistic electronic states. According to Eq.~3, the dipole term ($\propto\lambda_{\rm Dip}$) together with the SOC term cause the spin polarized in the $z$ direction while the vertical-field terms ($\propto\lambda_{\rm V}'$ and $\lambda_{\rm V}''$) give rise to an in-plane spin polarization (see the Supplementary Information for detailed discussions). As a result, a canted spin texture is expected for the Weyl cone. The two valleys of Weyl fermion states possess opposite spin polarizations because the two valleys are partners under the time-reversal symmetry. This is indeed what we found in the first-principles calculations and spin-resolved ARPES measurements as shown in Fig.~4. The calculated spin polarization of bands along ``cut1" and ``cut2" (marked in Fig.~4{\bf h}) are shown in Figs.~4{\bf a-f}. Along ``cut1", the Weyl bands near the fermi level carries nonzero $s_z$ and $s_y$ components. The absence of $s_x$ component is due to the fact that the $\overline{{\Gamma}}-\overline{\textrm{X}}_{1}$ direction corresponds to a glide line of the lattice (see Supplementary Information). By contrast, all three spin components show up in the bands along ''cut2". The two linear bands of the Weyl cone in ``cut2" have the same sign in the $s_{y,z}$ components but opposite signs in the $s_{x}$ component. The spin orientation of the states ``A1"-``A4" at the lower Weyl cone is schematically plotted in Figs.~4{\bf g,h}, which demonstrates the canted spin texture of the Weyl fermion states. To verify the spin texture of the Weyl fermion states, we performed spin-resolved ARPES measurements on bismuthene/SnSe. Spin-resolved momentum distribution curves (MDC) taken at $E=-0.2$ eV below the Weyl node (along a line marked by the dashed arrow in Fig.~4{\bf d}) are shown in Figs.~4{\bf i-k}. The blue and red dotted lines are photoemission intensity recorded in the ``spin-up" and ``spin-down" channels for the corresponding spin component, respectively. The spin polarization (defined as $P=\frac{1}{S_\textrm{eff}}\frac{I_{+}-I_{-}}{I_{+}+I_{-}}$, where the effective Sherman function $S_\textrm{eff}$ = 0.275 for our spin detectors) is shown in Figs.~4{\bf l-n}{\cite{Rev.Sci.Instrum.70.3572(1999), Rev.Sci.Instrum.79.123117(2008)}. The observed spin polarization of $\langle s_{x}\rangle$ and $\langle s_{y}\rangle$ is in good agreement with the theoretical results. The observed $\langle s_{z}\rangle$ is less than the calculated value, which can be attributed to the existence of rotated Bi domains (see Supplementary Information). Nonetheless, the two branches of the Weyl cone  have the same sign in the observed $\langle s_{z}\rangle$, which is consistent with the theoretical result in Fig.~4{\bf f}. The measured spin polarization of states from the other valley along cut 3 shown in Figs.~4{\bf o,p} provides a full view of the spin texture of the 2D Weyl semimetal. Our spin-ARPES results unambiguously confirm the canted spin texture and linear band dispersion of the Weyl cone in the epitaxial bismuthene.

\section{Bulk-boundary correspondence}
 To show the unique bulk-boundary correspondence of 2D Weyl semimetals, we calculated the bands of a semi-infinite bismuthene/SnSe heterostructure with an open boundary in the (010) direction. The result is plotted in Fig.~5{\bf a}. The Fermi string edge band directly connect the two bulk Weyl nodes as required by the band topology of the 2D Weyl cone. This Fermi string band is topologically protected and thus is robust against perturbations as shown in Supplementary Information. The connection of the ESBs to the bulk bands is schematically shown in Fig.~5{\bf b}. Besides the Fermi string band, extra in-gap edge state bands exist between the two Weyl nodes. The existence of those extra edge state bands can be attributed to the fact that the 2D Weyl semimetal is at a critical point in connection to two topologically distinct insulator phases (see Supplementary Information for detailed discussion). The Fermi string edge band gives rise to an enhanced local density of states (LDOS) at the edge, especially, in a narrow energy window around the energy of bulk Weyl nodes. The LDOS can be directly probed by the differential conductivity dI/dV spectrum in scanning tunneling spectroscopy (STM) experiments. An STM topography of bismuthene on SnSe is shown in Fig.~5{\bf c}. The surface of SnSe, the interior of bismuthene, and the edge of bismuthene are marked by the black, green, and magenta dots, respectively. We measured the averaged dI/dV spectrum at 4.6K from the three regions (black, green, and magenta), and the result is plotted in Fig.~5{\bf d}. A large gap of$\sim$1.3~eV is observed in SnSe, indicating the bulk Weyl states are entirely confined with the Bi overlayer. The spectral curve from the interior of bismuthene demonstrates a vanishing LDOS at the energy of Weyl nodes, $E_\textrm{W}$, and linearity near $E_\textrm{W}$, which is consistent with the linear band dispersion of Weyl cones. Remarkably, the edge dI/dV spectrum (magenta) shows a higher LDOS compared to the bulk spectrum (green) in a narrow energy window around $E_\textrm{W}$ marked by the two blue arrows. This enhanced LDOS can also be seen in the dI/dV maps taken near an edge (Fig.~5{\bf e}). The edge is brighter than surfaces of SnSe and bismuthene at $V=-12$ and 17~meV. A similar edge spectrum was observed in bismuthene/SnS samples (see Supplementary Information). The results are in accordance with the existence of topologically protected edge states near the Fermi level. Interestingly, the high-resolution dI/dV map reveals a unique plane wave-like pattern near the energy of Weyl nodes as shown in Fig.~5{\bf f}. The Fourier transform of the dI/dV map gives the quasiparticle interference pattern (Fig.~5{\bf g}), in which there is an oval contour at the center of the $\bm{q}$ space accompanied by two small satellite contours. The distance between the satellite and the center of the oval is $\Delta q=0.42$~\AA$^{-1}$, which is exactly the separation between the two Weyl nodes in the 2D Brillouin as measured from ARPES results. Thus, the central oval contour in the QPI is from intravalley scatterings while the two satellites are induced by intervalley scatterings. The experimental QPI is consistent with the calculated result shown in Fig.~5{h}. The plane wave-like QPI with a single wavevector reflects the nodal Fermi surface of the 2D Weyl semimetal in epitaxial bismuthene.  

\section{Conclusion}
 The observation of gapless linear dispersion and spin texture of Weyl bands together with the enhanced edge density of states demonstrate epitaxial bismuthene/SnS(Se) as an ideal 2D Weyl semimetal with topological Fermi string edge states. This finding completes the family of solid-state Dirac and Weyl semimetals. The 2D Weyl semimetal in epitaxial bismuthene on SnS(Se) provides unique opportunities for exploring exotic relativistic and topological phenomena pertaining to Weyl fermions in 2D space.

\section{Methods}
\subsection{Growth of bismuthene on SnS and SnSe}
Bi was deposited on the cleaved surface of SnS and SnSe crystals in an MBE-ARPES-STM ultrahigh vacuum (UHV) system. The SnS and SnSe crystals are n-type doped with Br. The base pressure was lower than $2 \times 10^{-10}$~mbar. High-purity Bi was evaporated from a standard Knudsen cell with a flux of 0.3~\AA/min. The temperature of the substrate was kept at $50^\circ \textrm{C}$ during the growth. The substrate temperature is critical for growing smooth Bi monolayer in the phosphorene structure.

\subsection{Scanning tunneling microscopy measurement}
An {\it in-situ} Aarhus-150 STM was used to characterize the surface topography and the lattice parameters of the $\-$-Sb films. The topography was measured at room temperature with the base pressure lower than $2 \times 10^{-10}$~mbar. The bias voltage and the tunneling current were set to be 1.5~V and 0.01~nA for the surface topography measurement,  5~mV and 0.15~nA for the zoom-in atom-resolved STM measurement. The dI/dV spectrum, dI/dV mapping, and the STM QPI were produced by an Omicron LT-Nanoprobe system at 4.6K. The sample was transferred through a ultra-high vacuum (UHV) suitcase with pressure below $1 \times 10^{-9}$~mbar. The tunneling current is set to be 200~pA during the dI/dV measurement. 

\subsection{Spin- and Angle-resolved photoemission spectroscopy measurements}
Spin and angle resolved photoemission spectroscopy measurements were performed in a lab-based system coupled to the molecular beam epitaxy system, using a Scienta DA30L hemispherical analyzer with a base pressure of $P<$ 5$\times$10$^{-11}$ mbar and a base temperature of $T\sim$8 K.  Samples were illuminated with linearly polarized light using an Oxide $h\nu =$11 eV laser system. The light polarization was set perpendicular to both the sample and the slits of the detector. For electronic dispersion measurements, a pass energy of 2 eV and 0.3 mm slit was used for a total energy resolution $\sim$2.5 meV and momentum resolution $\sim$0.01 $\mathrm{\AA}^{-1}$.  Dual VLEED ferrums that utilize exchange scattering are coupled to the electron analyzer and used to determine the spin $s_x$, $s_y$, and $s_z$ polarizations of the measured electrons.  For spin resolved measurements a pass energy of 10 eV and a 1 mm $\times$ 2 mm spin aperture was used yielding a total energy resolution $\sim$50 meV and momentum resolution $\sim$0.033 $\mathrm{\AA}^{-1}$.

\subsection{First-principles calculation}
First-principles calculations with Density Function Theory (DFT) were performed by using the Vienna ab Initio Simulation Package (VASP) package \cite{PhysRevB.54.11169}. The Perdew-Burke-Ernzerhof (PBE) \cite{PhysRevLett.77.3865} exchange-correlation functional was used. The experimental lattice parameter was applied for bismuthene; the lattice parameter of SnSe(S) was modified to match the covered Bi. The spin-orbit coupling (SOC) was included self-consistently in the calculations of electronic structures with a Monkhorst-Pack 11$\times$11$\times$1 $k$-point mesh. The vacuum thickness was greater than 20~\AA \ to ensure the separation of the slabs. Atomic relaxation was used until the residual forces were less than 0.01~eV/\AA. 

We constructed a tight-binding Hamiltonian for Bi/SnSe(S), where the tight-binding model matrix elements were calculated by projecting onto the Wannier orbitals\cite{PhysRevB.56.12847, PhysRevB.65.035109, MOSTOFI2008685}, which used the VASP2WANNIER90 interface. The Bi $p$ orbitals, Sn $p$ orbitals, and Se $p$ orbitals were used to construct the Wannier functions without performing the maximizing localization. The edge state electronic structure was calculated by the Green’s function technique, which computes the spectral weight near the edge of a semi-infinite system. To simulate the photoemission matrix element effects in the ARPES spectra, we consider a non-trivial structure factor in ab initio calculations \cite{Lee_2018, PhysRevX.7.041067}. To simulate this effect, we construct a unitary matrix $U(\textbf{\textit{k}})$:
\begin{equation*}
U(\textbf{\textit{k}}) = 
\begin{pmatrix}
e^{i\textbf{\textit{k}}\cdot\textbf{\textit{r}}_1} & \cdots & 0 \\
\vdots  & \ddots & \vdots  \\
0 & \cdots & e^{i\textbf{\textit{k}}\cdot\textbf{\textit{r}}_n} 
\end{pmatrix},
\end{equation*}
where $\textbf{\textit{k}}$ is the momentum vector, $\textbf{\textit{r}}_i$ is the real space coordinates of $i$th atom in the original Bi/SnSe(S) unit cell. We simulate the unfolding band structures of Bi/SnSe via applying this unitary matrix to the tight-binding Hamiltonian, $U(\textbf{\textit{k}})H(\textbf{\textit{k}})U(\textbf{\textit{k}})^\dagger$. 

 The quasiparticle interference pattern was calculated based on the Green’s function method by using spin-dependent scattering probability (SSP) method \cite{PhysRevB.93.041109, science.aad8766}, which can be written as 
\begin{equation*}
J_{s}(q)=\frac{1}{2} \sum_k \sum_{i=0,1,2,3} \rho_{i} (k) \rho_{i} (k+q) ,
\end{equation*}
where $\rho_{0}(k)=\textrm{Tr}[G(k)]$ is the total spectral density and $\rho_{i}(k)=\textrm{Tr}[\sigma_{i} G(k)]$ is the spin density, in which $G(k)=[\omega+i\eta-H(k)]^{-1}$ is Green’s function of the system and $\sigma_{i=1,2,3}$ are the Pauli matrices for spin.

\bibliographystyle{ieeetr}
\bibliography{2DWeyl}

\newpage

\begin{figure}
\includegraphics[width=1.0\linewidth]{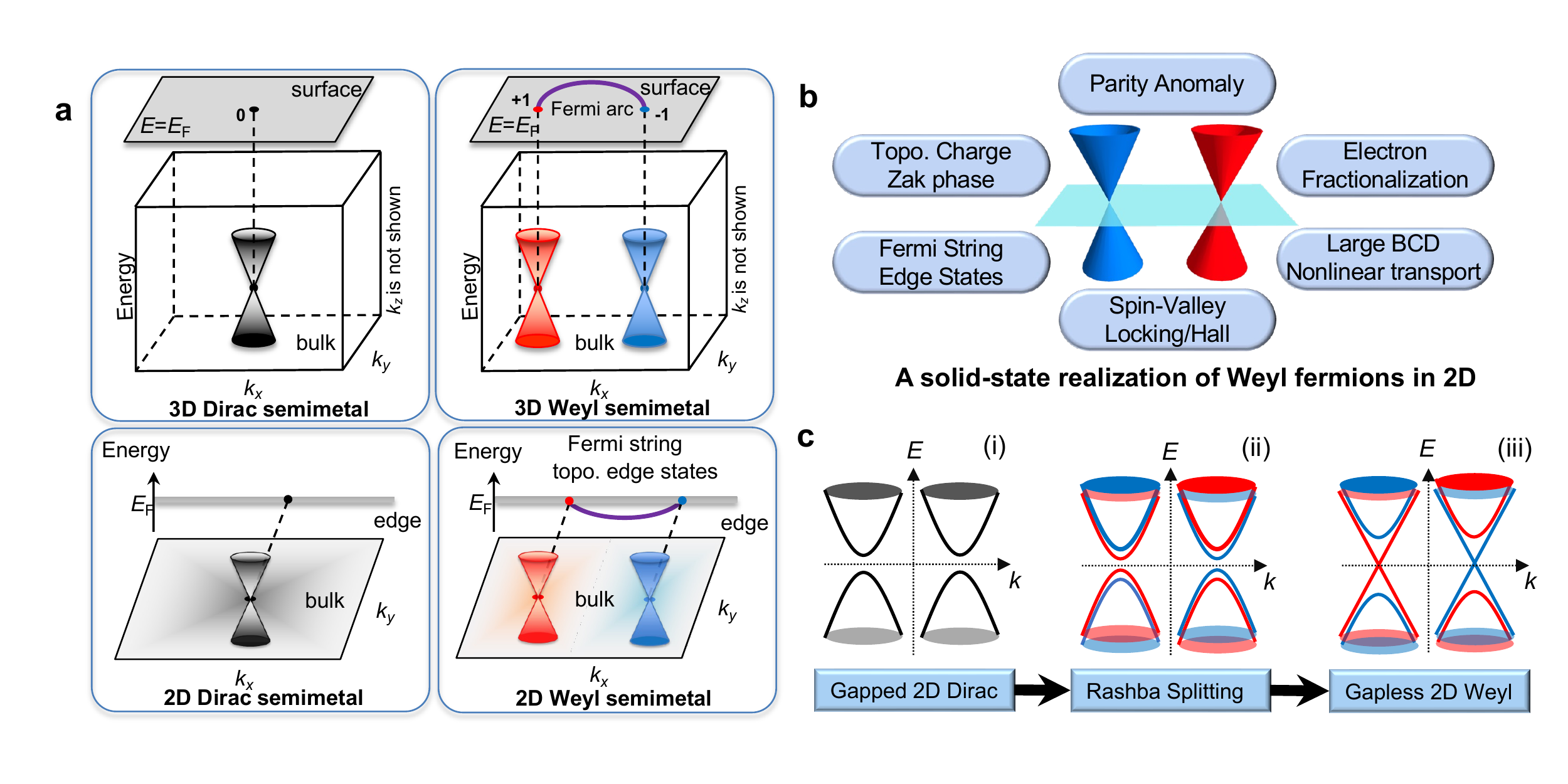}
\caption{\sf{\bf\sffamily A new topological phase of matter: 2D Weyl semimetals.} {\bf\sffamily a}  Overview of Dirac/Weyl semimetals and their topological boundary states. {\bf\sffamily b} Summary of exotic properties of 2D Weyl semimetals. {\bf\sffamily c} Schematic formation mechanism of 2D Weyl semimetals in epitaxial bismuthene. The red and blue colors indicate two opposite directions of spin polarization.}%
\end{figure}

\newpage

\begin{figure}
\includegraphics[width=1.0\linewidth]{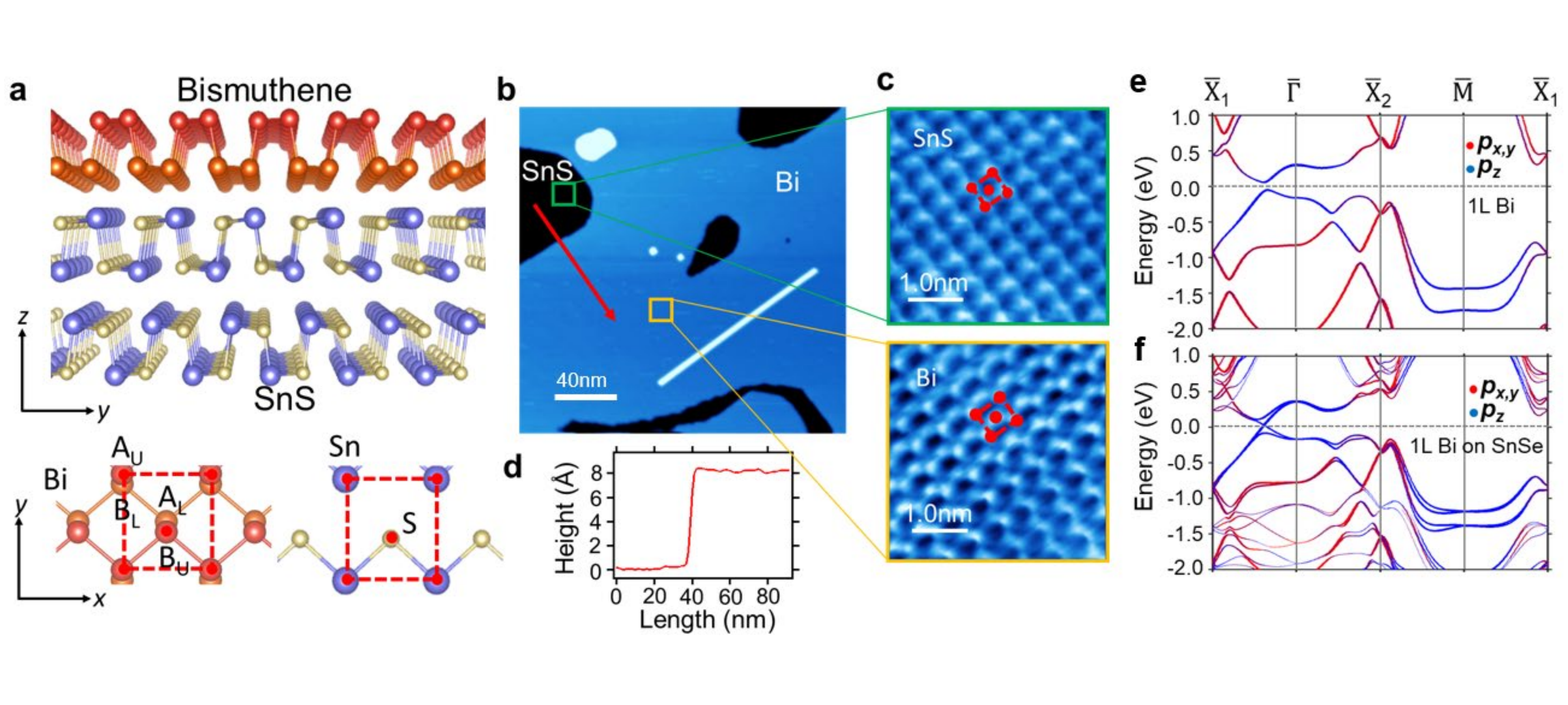}
\caption{\sf{\bf\sffamily Lattice property, band structure and STM characterization of bismuthene grown on SnS.} {\bf\sffamily a} Side and top views of bismuthene lattice structure. {\bf\sffamily b} Large-scale STM image of bismuthene grown on SnS substrate. {\bf\sffamily c} Zoom-in STM images of bismuthene and the surface of SnS. {\bf\sffamily d} The height profiles taken along the red arrow in {\bf\sffamily b}. {\bf\sffamily d}  Calculated band structure of free-standing bismuthene (top) and epitaxial bismuthene on SnSe (bottom). 
}%
\end{figure}

\newpage
\begin{figure}
\includegraphics[width=0.9\linewidth]{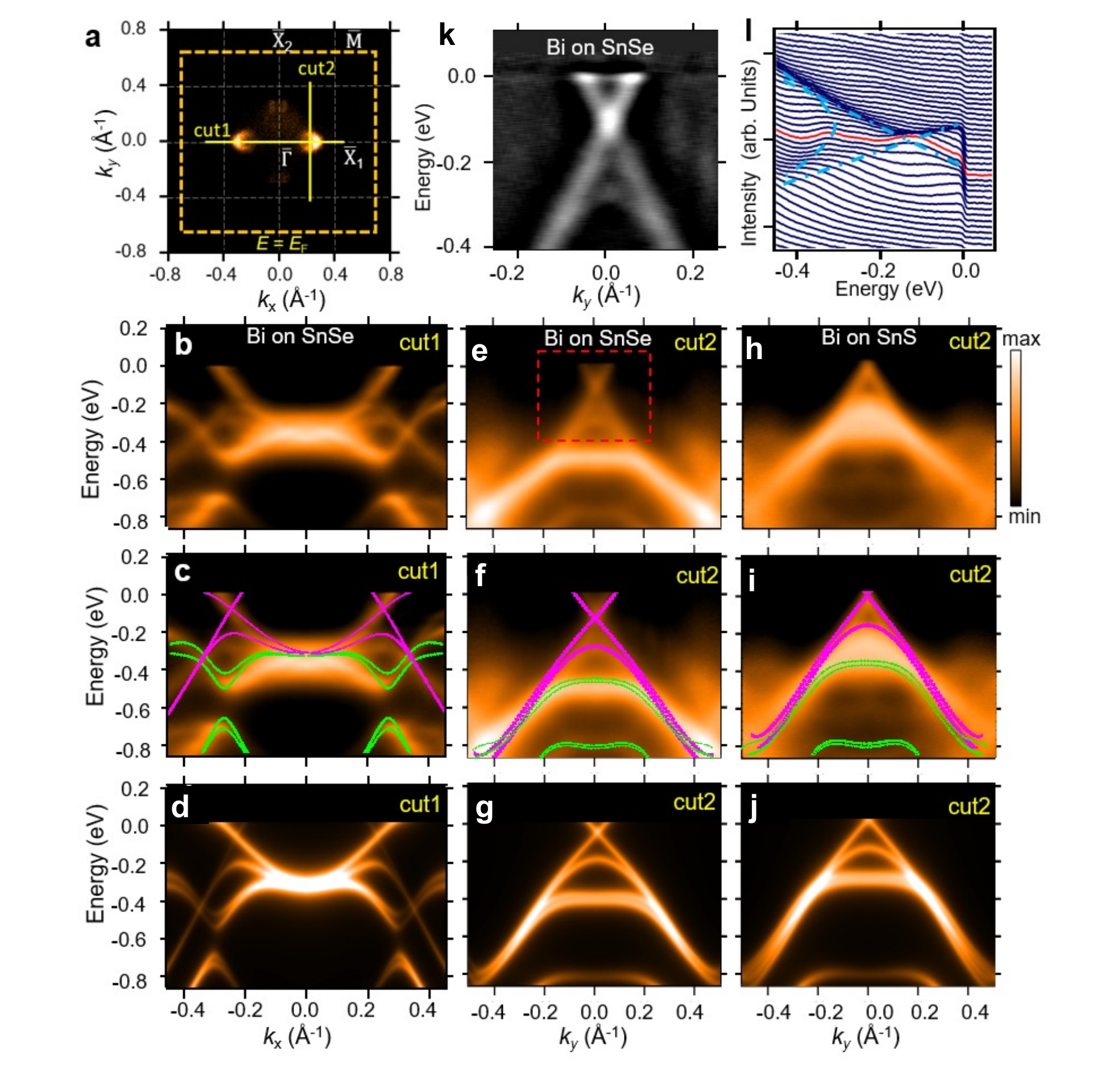}
\caption{\sf{\bf\sffamily First-principles band structure and ARPES spectra of epitaxial bismuthene. } {\bf\sffamily a} ARPES Fermi surface taken from bismuthene on SnSe. {\bf\sffamily b} ARPES spectra of bismuthene taken along the line of ``cut1" marked in {\bf\sffamily a}. {\bf\sffamily c} Overlay of calculated band structure on the ARPES spectrum along ``cut1". The magenta lines are bands along the direction of $\overline{\Gamma}$-$\overline{\textrm{X}}_1$ while the green lines are bands in the direction perpendicular to $\overline{\Gamma}$-$\overline{\textrm{X}}_1$.  {\bf\sffamily d} Calculated spectra with the inclusion of photoemission matrix elements. {\bf\sffamily e-g} Same as {\bf\sffamily b-d} but for bands of bismuthene on SnSe along ``cut2". {\bf\sffamily h-j} Same as {\bf\sffamily b-d} but for bands of bismuthene on SnS along ``cut2". {\bf\sffamily k} Second derivative of the ARPES spectrum in the red box in {\bf\sffamily e}. {\bf\sffamily l} Map of energy distribution curves (EDC) for the ARPES spectrum in the red box in {\bf\sffamily e}. The blue dotted lines mark the maximum of each EDC. The red solid line plot the EDC at the momentum of the Weyl point.
}%
\end{figure}

\newpage
\begin{figure}
\includegraphics[width=1.0\linewidth]{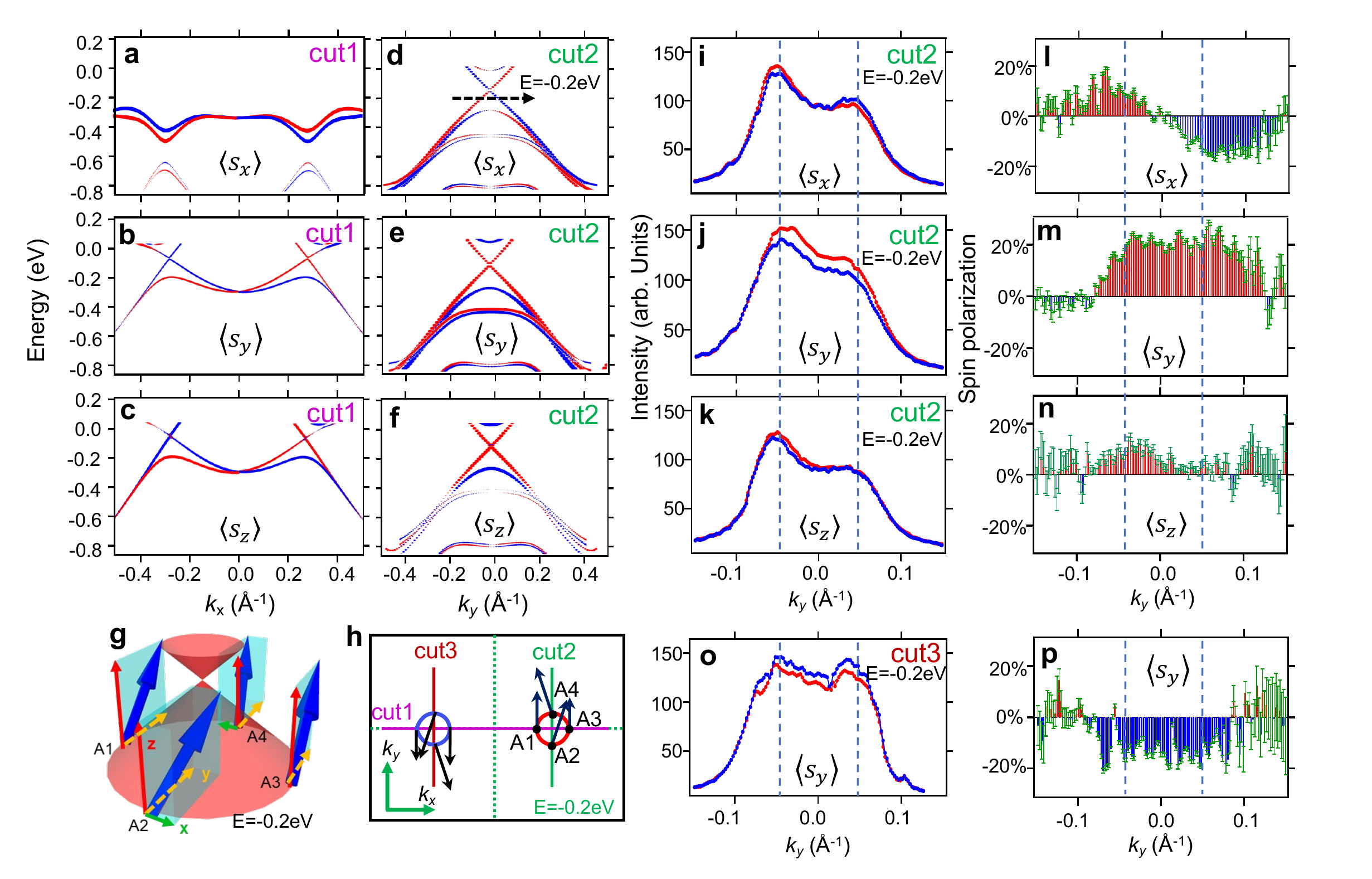}
\caption{\sf{\bf\sffamily Spin texture of 2D Weyl cones in bismuthene on SnSe.}  {\bf\sffamily a-c} The calculated spin components $\langle s_{x}\rangle$, $\langle s_{y}\rangle$, and $\langle s_{z}\rangle$ of the bismuthene bands along ``cut1" marked in {\bf\sffamily h}, respectively. The blue and red colors represent the ``spin-down" and ``spin-up" states of the corresponding spin component. {\bf\sffamily d-f} Same as {\bf\sffamily a-c}, but for the line of ``cut2" marked in {\bf\sffamily h}. {\bf\sffamily g} Spin orientation of states A1-A4 marked in {\bf\sffamily h}. The length of the arrow indicates the magnitude of spin polarization. {\bf\sffamily g} In-plane spin texture of iso-energy contours at $E=-0.2$~eV. {\bf\sffamily i-k} Spin-resolved momentum distribution curves (MDC) taken at $E=-0.2$~eV along the line marked by the black dashed arrow in {\bf\sffamily d}. The blue and red curves are photoemission intensity recorded in the ``spin-down" and ``spin-up" channels, respectively, for $\langle s_{x}\rangle$, $\langle s_{y}\rangle$, and $\langle s_{z}\rangle$. {\bf\sffamily l-m} The shaded area with error bars indicate net spin polarization of $\langle s_{x}\rangle$, $\langle s_{y}\rangle$, and $\langle s_{z}\rangle$, respectively, which is calculated based on the spin results in {\bf\sffamily i-k}. {\bf\sffamily o,p} Spin-resolved MDC and calculated spin polarization of $\langle s_{y}\rangle$ at $E=-0.2$~eV along ``cut3" from the other valley. 
}%
\end{figure}

\newpage
\begin{figure}
\includegraphics[width=1.0\linewidth]{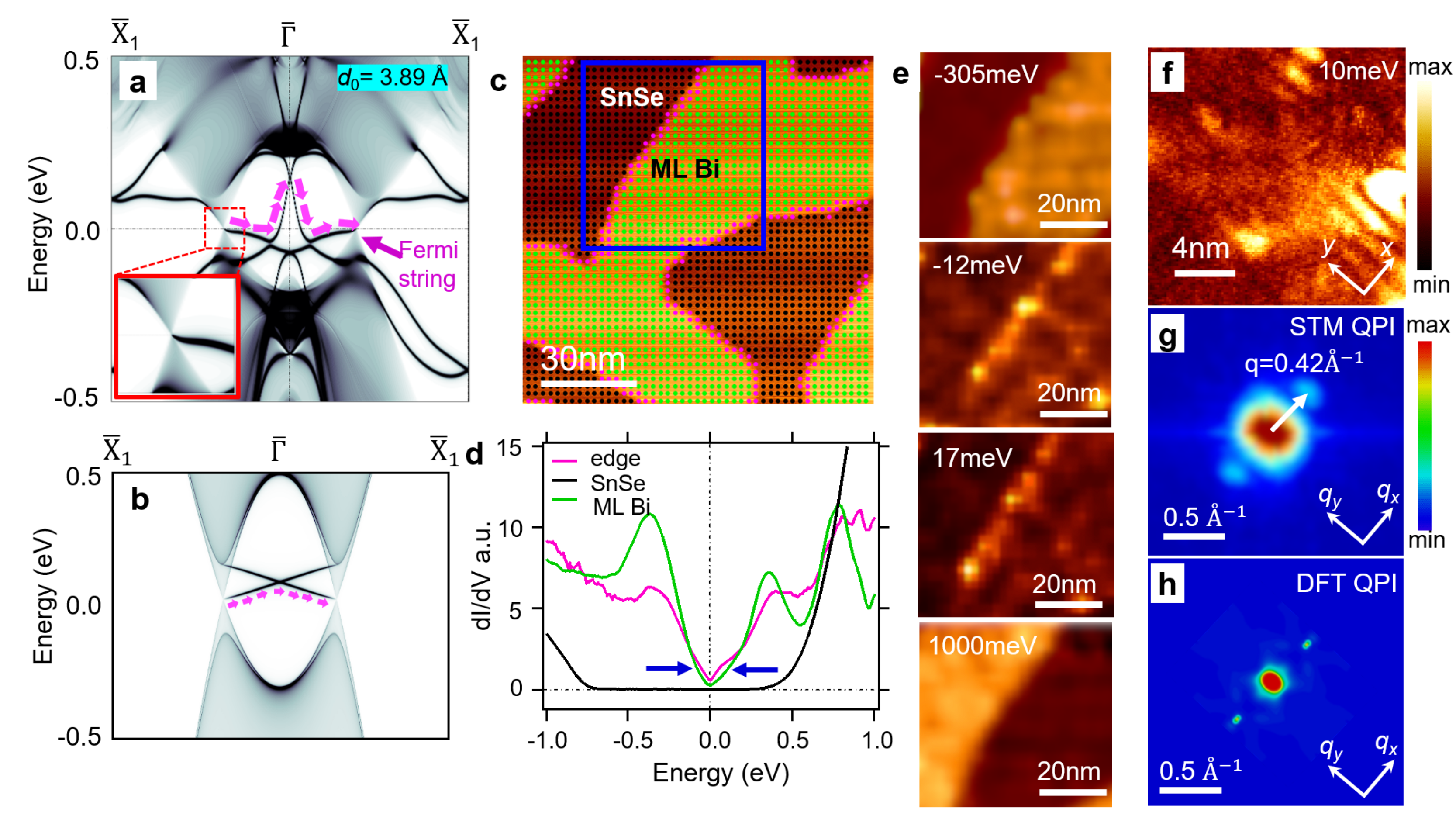}
\caption{\sf{\bf\sffamily Bulk-boundary correspondence in 2D Weyl semimetals.} {\bf\sffamily a} The edge bands and projected bulk bands of a semi-infinite bismuthene film on SnSe with an open boundary in the (010) direction. The bands are weighted with the charge density near the edge. {\bf\sffamily b} Schematic of the connection of edge bands to bulk Weyl bands. The magenta dashed curve plots the topological Fermi string edge band connecting the two Weyl nodes. {\bf\sffamily c} STM topography of bismuthene on SnSe. {\bf\sffamily d} Differential conductivity dI/dV spectra (aligned with the energy of Weyl nodes, $E_\textrm{W}$) taken at different locations. The green curve is the averaged spectrum taken at the green grid inside the bismuthene patch shown in {\bf\sffamily c}. The black curve is the averaged spectrum from the black grid on the surface of SnSe. A large gap of $\sim$ 1.3 eV is observed in SnSe. The magenta curve is the averaged spectrum from the magenta points at the edge of the bismuthene patch. The energy window with higher LDOS at the edge is marked by the two blue arrows. {\bf\sffamily e} dI/dV map over the area marked by the blue box in {\bf\sffamily c} at bias voltage $V$= $-305$, $-12$, $17$, and $1000$~meV. {\bf\sffamily f} High-resolution dI / dV map at bias voltage $V$= $+10$~meV. {\bf\sffamily g} The corresponding Fourier transform of the dI/dV map in {\bf\sffamily f} showing the quasiparticle interference pattern. {\bf\sffamily h} The quasiparticle interference pattern calculated by using spin-dependent scattering probability method. 
}
\end{figure}

\end{document}